\documentclass[preprint,aps,12pt,showpacs,nofootinbib,tightenlines]{revtex4}
\usepackage{graphicx}
\usepackage{booktabs}
\usepackage{amssymb,bm,mathrsfs,bbm,amscd}
\usepackage[tbtags]{amsmath}
\usepackage{lastpage}
\usepackage{CJK}
\usepackage{amsmath}
\usepackage{amssymb}
\usepackage{epsfig}
\usepackage{graphicx}
\begin{document}
\def\pslash{\rlap{\hspace{0.02cm}/}{p}}
\def\eslash{\rlap{\hspace{0.02cm}/}{e}}
\title {Production and decay of for the 125 GeV Higgs boson in the littlest Higgs model with T-parity  }

\author{%
Qing-Guo Zeng  $^{1,2}$}
 \author{ Shuo  Yang $^{3,4}$}
\author{ Chong-Xing Yue$^{1}$ }\email{cxyue@lnnu.edu.cn}
\author{  Lian-Song Chen$^{1}$}

\address{%
$^1$ Department of Physics, Liaoning Normal University, Dalian 116029,  China\\
$^2$ Department of  Physics, Shangqiu Normal University, Shangqiu  476000,  China\\
$^3$ Physics Department, Dalian University, Dalian, 116622,  China\\
$^4$ Center for High Energy Physics, Peking University, Beijing, 100871,  China
 \vspace*{1.5cm}}

\begin{abstract}

  Motivated by recent search results for the standard model (SM) Higgs
boson at the Large Hadron Collider (LHC), we revisit the Higgs
phenomenology in the littlest Higgs  model with T-parity  (LHT). We
present the signal strength modifier $\mu$ respectively for the main
search channels $qq' \rightarrow jjh\rightarrow jj\gamma\gamma$,
   $qq'\rightarrow Vh\rightarrow V\gamma \gamma$, $qq'\rightarrow Vh \rightarrow Vbb$, $gg\rightarrow h \rightarrow \gamma\gamma$,
   and $gg \rightarrow h \rightarrow VV$ in the  LHT  model. It is found that an enhancement factor of $1.09-1.56$
   in $qq' \rightarrow jjh\rightarrow jj\gamma\gamma$ channel can be obtained for this model
   in Case B with parameter $f$ in the range $1000$ GeV$\sim 500$ GeV. However, the rates for $b\bar{b}$, $\tau\bar{\tau}$
   are significantly suppressed relative to the SM predictions which are still
   consistent with the current sensitivity. It is hoped that will be further tested with larger integrated luminosity at the LHC.

\end{abstract}
\pacs{14.60.Hi, 12.60.-i, 13.66.De} \maketitle
\section{ Introduction}
The standard model (SM) is built on two cornerstones. One is gauge
theory which has been confirmed by the discovery of the electroweak
gauge bosons W and Z, and is further verified by electroweak precision
measurements. However, the other one, the electroweak symmetry
breaking (EWSB) mechanism,  still requires a  direct  experimental test.
The Large Hadron Collider (LHC) bears the responsibility for
detecting the Higgs boson and revealing the mystery of EWSB.

Over a long period, the Higgs mass is a free parameter in a wide range
from 114.4  GeV to 700 GeV which is based on the LEP search bound
\cite{LEP} and unitary constraint \cite{unitary}. Lately, the Tevatron
has excluded the production of a Higgs boson in the narrow mass
window $156\sim177$ GeV at 95\% CL \cite{Tevatron}. The LHC
experiment has the potential to cover the Higgs boson search in the
mass range from 100 GeV to TeV order. For a light Higgs with a mass
below 120 GeV, the best search channel is $h\rightarrow \gamma
\gamma$. At a medium mass range of $120-200$ GeV, the best sensitivity is
achieved in the $WW$ channel. With the increase in the Higgs mass,
the search channel $h\rightarrow ZZ\rightarrow 4l$ and $h\rightarrow
ZZ\rightarrow 2l2\nu$ becomes more sensitive.

 Recently,  the ATLAS and CMS Collaborations have
released the search results for the SM Higgs based on the 2012 data
corresponding to a luminosity of about $5\sim 6$ fb$^{-1}$ at 8 TeV
and the 2011 data around $5$ fb$^{-1}$ at the 7 TeV run \cite{ATLAS,CMS}.
Both Collaborations further confirmed the previous event excess
\cite{ATLASold,CMSold} and announced the discovery of a Higgs-like
particle around 125 GeV with a local significance of the $5\sigma$
level.

It is interesting to note that the present Higgs search results from ATLAS and CMS still hint at
 a larger observed rate in $h\rightarrow \gamma \gamma$ than the SM prediction although
 this is not much evident as the previous results \cite{ATLAS,CMS,ATLASrr,CMSVBF}.
 Motivated by recent search results especially the $h\rightarrow \gamma \gamma$ excess,
extensive studies have been carried out including attempts to draw more information from
the data and to use a global fit to get the best constraints on the Higgs effective couplings \cite{Hfit1,Hfit2,Hfit3}.
The global fit results show that the SM Higgs boson with a mass around 125 GeV can correctly predict
 the observed rates but better fits are obtained by some non-standard scenarios
 that predict more $\gamma \gamma$ signal events \cite{Hfit1,Hfit2,Hfit3}.

In many new physics models, new particles are introduced and they
will contribute significantly to  Higgs production and decay. The
current Higgs search results could provide clues into the underlying
physics and then generate profound effects on new physics searchs.
After the discovery of the Higgs-like boson, the next important
mission is to test the properties of the particle including its
couplings to the SM particles and its spin and CP quantum numbers.
Extensive studies on Higgs phenomena in different models are needed
to verify, constrain or rule out different new physics models
according to the data.

In the little Higgs models, new particles coupling to the Higgs boson are
introduced to cancel the quadratic divergence of the Higgs mass
induced by SM particles. Although  no direct signal of little Higgs
particles has been found until now, the search results for the Higgs
boson can give important indirect constraints on these new particles
and the model parameter space. The Higgs phenomenology in the
littlest Higgs model with T-parity (LHT) \cite{LHT} has been
extensively studied in Refs. \cite{LHTrr,LHTrr2,LHTrr3}.
 In this paper, we revisit the Higgs boson in the LHT model  and test the parameter space in
light of recent search data. We present the total width and the
branching ratios for the Higgs boson with 125 GeV mass in the LHT
model. It is expected that the width information of the Higgs boson
in the LHT  model could be further tested with a larger data sample.
Furthermore, we study the search channels $qq' \rightarrow jjh
\rightarrow jj\gamma \gamma$, $qq'\rightarrow Vh$ followed by
$h\rightarrow b\bar{b}$, $gg\rightarrow h\rightarrow \gamma \gamma$,
$gg\rightarrow h \rightarrow WW$, and $gg\rightarrow h\rightarrow
ZZ$.

The rest of this paper is organized as follows. In Section 2, we
 briefly review the LHT model. In Section 3, we
calculate the width information of the 125 GeV Higgs boson in the
LHT model, and we also calculate the rates of the main Higgs search
channels normalized to the SM prediction in the LHT model. Finally,
we  give our conclusions in Section 4.
\section{ The Higgs in the LHT model }
 \noindent

A key feature of the little Higgs theory is that the Higgs boson
is a pseudo-Goldstone boson from the breaking of a large global
symmetry and the EWSB is triggered by the Coleman-Weinberg potential
\cite{LH}. In this paper,
 we shall focus on  the LHT model  which
 has been the most popular little Higgs model in recent years. In the LHT model,
 a discrete parity called T-parity is introduced into the littlest Higgs model \cite{LHM}
  and particle fields are divided into T-even and T-odd sectors under the parity and the SM fields are T-even.
    Because all the dangerous tree-level contributions to
low energy EW observables are forbidden by T-parity,
the relatively low symmetry breaking  scale $f$ is allowed.

To implement T-parity, two fermion $SU(2)$ doublets, $q_{1}$ and
$q_{2}$ as $q_{i}=-\sigma_{2} (u_{L_{i}},
d_{L_{i}})^{T}=-(id_{L_{i}}, -iu_{L_{i}})^{T}$ with $i = 1$ and $2$,
are introduced for each SM fermion doublet.
 $q_{1}$ and $q_{2}$ are embedded into incomplete $SU(5)$ multiplets
$\Psi_{1}$ and $\Psi_{2}$ as $\Psi_{1} = (q_{1}, 0, 0_{2})^{T}$ and $\Psi_{2}  = (0_{2}, 0, q_{2})^{T}$,
 where $ 0_{2} = (0, 0)^{T}$. A multiplet $\Psi_{c}$ is introduced as $\Psi_{c} = (q_{c}, \chi_{c}, \tilde{q}_{c})^{T}$.

The fermion mass terms and interaction terms
with the neutral Higgs boson are given by \cite{LHT}
\begin{eqnarray}
\label{eq2}
{\cal L}_{\kappa} &=& -\sqrt{2} \kappa f[
\bar{d}_{L_-} \tilde{d}_{c}+\frac{1+c_\xi}{2} \bar{u}_{L_-}
\tilde{u}_c
-\frac{s_\xi}{\sqrt{2}}\bar{u}_{L_-} \chi_c\nonumber\\&&
-\frac{1-c_\xi}{2} \bar{u}_{L_-} u_c]+h.c. \ .
\end{eqnarray}
\begin{eqnarray}
\label{eq2}
{\cal L}_t &=& -\lambda_1 f \left(
\frac{s_\Sigma}{\sqrt{2}} \bar{u}_{L_+} u_R
+\frac{1+c_\Sigma}{2} \bar{U}_{L_+} u_R
\right)\nonumber\\[1mm]
&&
-\lambda_2 f \left(\bar{U}_{L_+} U_{R_+}+\bar{U}_{L_-} U_{R_-}
\right)+ h.c.\ .
\end{eqnarray}
Here, $c_\Sigma(\equiv \cos\frac{\sqrt{2}(v+h)}{f})$ and
$s_\Sigma(\equiv \sin\frac{\sqrt{2}(v+h)}{f})$  originate  from the
non-linear sigma model field $\Sigma$. $u_{L_-}$ and $u_{L_+}$ are
defined by $u_{L_\pm} = (u_{L_1}\mp u_{L_2})/\sqrt{2}$ and they
correspond to the T-odd and T-even eigenstates, respectively.

The T-odd combination of the doublets $q_1$ and $q_2$   obtain a
mass $\sqrt{2} \kappa f$  from ${\cal L}_\kappa$ (cf. Eq.~(1)).
There are three generations of T-odd particles, here we assume they
are degenerate. A T-odd Dirac Fermion $T'$ ($T'_L \equiv
U_{L_-},~T'_R \equiv U_{R_-}$) gets a mass $m_{T'}=\lambda_2 f$ (cf.
Eq.~(2)). Note that $T'$ does not have tree-level Higgs boson
interaction, and thus it does not contribute to the $gg$ fusion
process at the one-loop order. The heavy T-even partner $(T)$ of the top
quark with the mass $m_{T}= \frac{m_{t}}{\sqrt{x_{L}(1-x_{L})}}
\frac{f}{\nu}$ is responsible for canceling the quadratic divergence
to the Higgs mass induced by the top quark where
$x_{L}=\frac{\lambda^{2}_{1}}{\lambda^{2}_{1}+\lambda^{2}_{2}}$.

The   effective Lagrangian   for down-type quark Yukawa
couplings in this paper is given by
\begin{eqnarray}
\label{eq2} {\cal L}_d &=& \frac{i\lambda_d}{2\sqrt{2}} f
\epsilon_{ij}\epsilon_{xyz}[
({\bar{\Psi}^{\prime}}_{2})_{x}\Sigma_{iy}\Sigma_{jz}X\nonumber\\[1mm]&&
-({\bar{\Psi}^{\prime}}_{1}\Sigma_{0})_{x}{\tilde{\Sigma}_{iy}}{\tilde{\Sigma}_{jz}}\tilde{X}]d_{R},
\end{eqnarray}
 where  ${\bar{\Psi}^{\prime}}_{1}=(-\sigma_{2}q_{1}, 0, 0_{2})^{T}$ and
${\bar{\Psi}^{\prime}}_{2}=(0_{2}, 0, -\sigma_{2}q_{2})^{T}$. Here
$X$ transforms into $\tilde{X}$ under T-parity, and
$X=(\Sigma_{33})^{-\frac{1}{4}}$ (denoted as Case A)and
$X=(\Sigma^{\dag}_{33})^{\frac{1}{4}}$(denoted as Case B) are chosen
 for $X$. Here $\Sigma_{33}$ is
the $(3, 3)$ component of the non-linear sigma model field $\Sigma$.

In addition to new Higgs interactions introduced in the LHT
model, the interactions between the Higgs boson and the SM particles
 are also modified as \cite{LHTrr}:
\begin{eqnarray}
\label{eq2}
 \frac{ g_{hVV}}{g^{SM}_{hVV}} &\thickapprox&
 1-\frac{1}{4}\frac{\upsilon_{SM}^{2}}{f^{2}}-\frac{1}{32}\frac{\upsilon_{SM}^{4}}{f^{4}},~~(V=
 Z,W), \\
  \frac{ g_{hu\bar{u}}}{g^{SM}_{hu\bar{u}}} &\thickapprox& 1-\frac{3}{4}\frac{\upsilon_{SM}^{2}}{f^{2}}-\frac{5}{32}\frac{\upsilon_{SM}^{4}}{f^{4}} ~~~(u=u,~c),\\
  \frac{ g_{hd\bar{d}}}{g^{SM}_{hd\bar{d}}} &\thickapprox& 1-\frac{1}{4}\frac{\upsilon_{SM}^{2}}{f^{2}}+\frac{7}{32}\frac{\upsilon_{SM}^{4}}{f^{4}} ~~~(Case~~A),\\
   \frac{ g_{hd\bar{d}}}{g^{SM}_{hd\bar{d}}} &\thickapprox& 1-\frac{5}{4}\frac{\upsilon_{SM}^{2}}{f^{2}}-\frac{17}{32}\frac{\upsilon_{SM}^{4}}{f^{4}} ~~~(Case~~ B).
   \end{eqnarray}
The relation of lepton Yukawa couplings are the  same  as the down-type Yukawa couplings.

\section{ Numerical Results and Analysis }

In this paper, in stead of studies of all the search channels, we mainly concentrate on five channels: vector boson fusion (VBF) followed by di-photon decay, i.e. $qq' \rightarrow jjh \rightarrow jj\gamma \gamma$, $qq'\rightarrow Vh$ followed by $h\rightarrow b\bar{b}$,
$gg\rightarrow h\rightarrow \gamma \gamma$, $gg\rightarrow h \rightarrow WW$, and $gg\rightarrow h\rightarrow ZZ$.
This is because among all the search channels these five channels provide a crucial role due to the large event rates and good reconstructed resolution.
In the LHT model, the additional T-even top partner and the T-odd fermions
 contribute significantly to the processes $gg\rightarrow h$ and $h\rightarrow \gamma\gamma$ induced at loop level.
Modified couplings of the SM particles as shown in Eq.~4-7 also
affect the production rates and decay widths of $h$.

In our calculation, the $gg$ fusion process is calculated by private codes using the loop functions, and VBF and Vh production processes are calculated with Madgraph4 \cite{Madgraph} where the parton distribution function CTEQ6L \cite{CTEQ6L} is used with the renormalization scale $ \mu_{R}$ and factorization scale $\mu_{F} $ chosen to be $ \mu_{R}= \mu_{F} = m_{h}$. The Higgs decay are calculated with HDECAY \cite{Hdecay}.

In the LHT model, the loop-induced partial decay width of $h\rightarrow \gamma\gamma$\footnote {A detailed study of loop-induced decay in littlest Higgs model can be found in Ref. \cite{HanHdecay}.} can be represented as

\begin{eqnarray}
\label{eq2}
   \Gamma(H \to \gamma\gamma)=\frac{\sqrt{2} G_F \alpha^2 m_H^3}{256 \pi^3}
    \mid \frac{4}{3} F_{1/2} (\tau_t) y_t y_{G_F}\nonumber\\
    + \frac{4}{3} F_{1/2} (\tau_T) y_T+\frac{4}{3} F_{1/2}(\tau_{T'}) y_{T'}\nonumber \\
    + F_1 (\tau_{W_L}) y_{W_L} y_{G_F} + F_1 (\tau_{W_H}) y_{W_H}
     \mid^2.
\end{eqnarray}
where $y_i$ represents the corresponding Higgs coupling in the LHT model. $T$, $T'$ and $W_H$ represent, respectively,
the T-even top partner, T-odd fermions and heavy charged gauge bosons in the model.
Here, the dimensionless loop factors \cite{HHG} are
\begin{eqnarray}
    F_1 = 2 + 3 \tau + 3\tau (2-\tau) f(\tau),\quad\nonumber\\
    F_{1/2} = -2\tau [1 + (1-\tau)f(\tau)],\quad\nonumber\\
    F_0 = \tau [1 - \tau f(\tau)],~~~~~~~~~~~~~~~~
\end{eqnarray}
with
\begin{equation}
    f(\tau) = \left\{ \begin{array}{lr}
        [\sin^{-1}(1/\sqrt{\tau})]^2, & \tau \geq 1 \\
        -\frac{1}{4} [\ln(\eta_+/\eta_-) - i \pi]^2, & \, \tau < 1
        \end{array}  \right.
\end{equation}
and
\begin{equation}
    \tau_i = 4 M_i^2 / m_H^2, \qquad
    \eta_{\pm} = 1 \pm \sqrt{1-\tau}.
\end{equation}

For the  $h\rightarrow gg$  process, the contributions from the T-even
top partner and T-odd fermions significantly suppress its rate
because in little Higgs models, the quadratic contribution to the Higgs
mass from the top quark is canceled by the contribution from its partner
which is derived from the underlying collective symmetry breaking.
For the $h\rightarrow \gamma\gamma$  decay, the W boson contribution
dominates over the top contribution in the SM and they will
partially cancel. The contributions from additional fermions tend to cancel
the contributions from new gauge bosons,  and the W boson still gives
 a dominant contribution. Therefore, the partial decay width of $h\rightarrow
\gamma \gamma$ does not vary much. However, the branching ratio of
$h \rightarrow \gamma \gamma$ is enhanced significantly due to total
width suppression in Case B. For precision, we have included all the
new particle contributions in the loop. We also find that the decay
$h\rightarrow \gamma \gamma$ is not sensitive to the parameter $x_L$
and $\kappa$. So does it for $gg\rightarrow h$ production. In this
paper, we fix the parameter $x_L=0.5$ and $\kappa=1$.

\begin{table}
\caption{ \label{tab2}   Higgs branching ratios and total widths
in the  LHT model.($m_{h}=125GeV$)} \vspace{-3mm} \footnotesize
\begin{tabular*}{155mm}{@{\extracolsep{\fill}}cccccccccccc}\\
\hline
\toprule f/(GeV)(case) & $b\bar{b}$ & $\tau\bar{\tau}$   & $c\bar{c}$  & $s\bar{s}$ & $\mu\mu$  & gg   \\
\hline
800\hphantom{0}\hphantom{0}(A)  &0.5922       &0.6489E-01     &0.2688E-01     &0.4508E-03    & 0.2252E-03     & 0.6221E-01     \\
1000\hphantom{0}(A) &  0.5857  & 0.6419E-01    &0.2770E-01  & 0.4457E-03  &0.2227E-03  &  0.7065E-01    \\
800\hphantom{0}\hphantom{0}(B) &0.5461       & 0.5984E-01  &0.3084E-01    & 0.4162E-03   &0.2076E-03    & 0.7229E-01  \\
1000\hphantom{0}(B) & 0.5578  & 0.6112E-01    & 0.3012E-01 & 0.4246E-03   &0.2121E-03  & 0.7740E-01  \\
\hline\hline
\toprule f/(GeV)(case)  &$\gamma\gamma$ & z$\gamma$ &ww &zz  &$t\bar{t}$   &Total$\Gamma_{h}$/(GeV)\\
\hline
800\hphantom{0}\hphantom{0}(A) &0.2383E-02    & 0.1651E-02   &0.2215       &0.2712E-01  &0   &0.379332E-02\\
1000\hphantom{0}(A)   & 0.2341E-02 &0.1609E-02 &0.2197    &0.2689E-01  &0 &0.389284E-02 \\
800\hphantom{0}\hphantom{0}(B)   &0.2734E-02    &0.1894E-02    &0.2541        &0.3111E-01  &0   &0.330616E-02  \\
1000\hphantom{0}(B)  & 0.2546E-02  &0.1749E-02  &0.2389  &0.2925E-01   &0 &0.357995E-02\\
\hline
\bottomrule
\end{tabular*}
\end{table}

We first  present the values of the total width and branching ratios
for two typical values $f=800$ GeV and $f=1000$ GeV  in Table I, and
it is expected that the width information of the Higgs in the LHT
model could be further tested at the LHC. For the SM Higgs boson
with a light mass of 125 GeV, the $b\bar{b}$ channel is the dominant
decay channel and the total width of the SM Higgs boson is about
$4.03$ MeV \cite{smwidth}. When it comes to the Higgs boson in the
LHT model, the dominant channel is still the $h\rightarrow
b\bar{b}$. However, there are some differences.

 In both Case A and Case B, the total widths of the Higgs boson are
suppressed. In Case A, the main decay channels of $b\bar{b}$,
$\tau\bar{\tau}$, $VV$ are slightly suppressed in similar factors
because the corresponding couplings are suppressed as shown in
Eqs.~4-7. While in Case B, the $b$ and $\tau$ Yukawa couplings are
significantly suppressed. For the Higgs with mass 125 GeV, the
$b\bar{b}$ and $\tau \bar{\tau}$ channels are the dominant decay
channels. Hence, the suppression of the Yukawa couplings of $h b\bar{b}$
and $h \tau\bar{\tau}$ results in the evident reduction in the total
width of the Higgs boson. However, it is interesting to note that this
also leads to an enhancement of the branching ratio of $h
\rightarrow \gamma\gamma$. The suppression factor of the branching
ratios of $h\rightarrow b\bar{b}$ and $h \tau\bar{\tau}$ and the
enhancement factor of the branching ratio of $h\rightarrow \gamma\gamma$
can be read directly from the table.

Here, we also present the total decay width of the Higgs boson in
the  LHT model normalized to the SM width as a function of $f$ in
Fig.~1. As shown in   Fig.~1, in both Case A and Case B the total
widths are suppressed in the whole $f$ range.
 When $f$ increases, the decoupling effect appears and the ratio is close to one.
  The suppression factors are about $ 0.838 \sim 0.988$ (Case A), and
  $0.543 \sim 0.967$ (Case B) for $f$ in the range $500$ GeV $\sim 2000$ GeV, respectively.
   As stated above, the suppression of the total widths is derived from the suppressions of the Higgs couplings in the LHT model.
    In particular, the total widths are significantly suppressed in Case B since the suppression of the Yukawa couplings
    of $h b\bar{b}$  directly affects the dominant decay channel $h\rightarrow b\bar{b}$.
\begin{figure}[htbp]
\includegraphics[width=9cm]{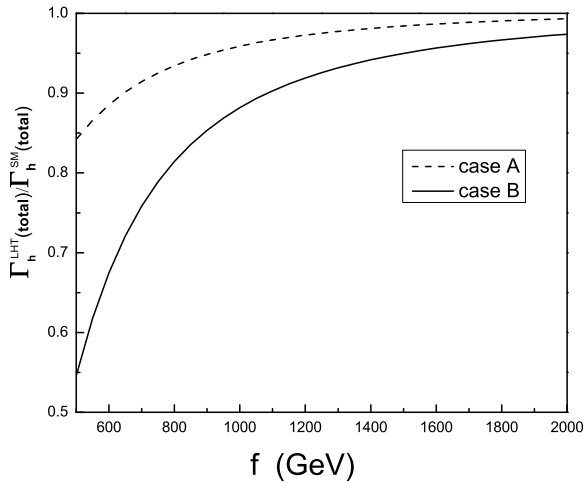}
\caption{\label{fig1}   $ \Gamma^{LHT}(total)/\Gamma^{SM}(total) $
as a function of  scale parameter $f$ .}
\end{figure}
Furthermore, we also calculate the hadron production cross sections
for the gluon-gluon fusion process $gg\rightarrow h$,
 the vector boson fusion process $qq' \rightarrow jjh$, the associated Higgs production with $W/Z$,
  $qq'\rightarrow Wh$ and $q\bar{q}\rightarrow Zh$ at the LHC. Both the two cases corresponding to
   $\sqrt{s}=7$ TeV and $\sqrt{s}=8$ TeV are considered and the results are shown in Fig.~2.
    Here, we mainly focus on the case for 125 GeV Higgs and our results are consistent with those results.
    As shown in Fig.~2, in the low $f$ range the cross section for gluon-gluon fusion
    is significantly suppressed as interpreted above and the other modes are also suppressed because of the modified couplings in the LHT model.
     When $f$ increases, the cross section is close to the SM prediction.
     From a phenomenological point, the difference between Case A and Case B is the coupling of $hd\bar{d}$ where d represents the down type fermions.
      So the cross sections shown in Fig.~2 are the same for Case A and Case B.
\begin{figure}[htbp]
\includegraphics [width=11cm]{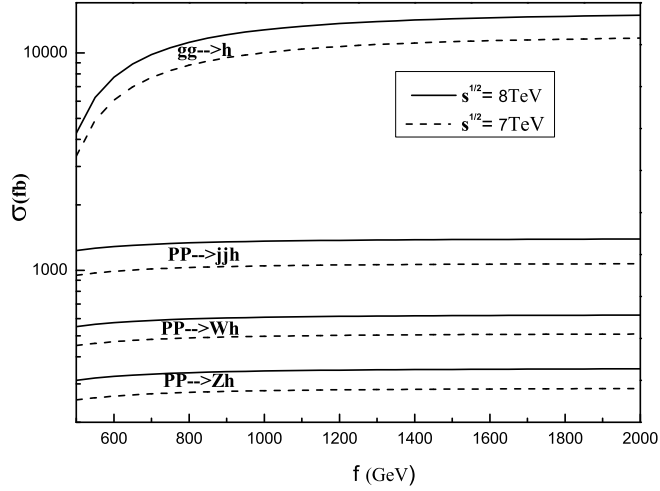}
\caption{\label{fig1}   The cross sections for the main modes of
Higgs production in the LHT model as  a function of  scale
parameter $f$ .}
\end{figure}
Here, we further consider the signal strength modifier $\mu_i=
\frac{\sigma(LHT)\times Br_i(LHT)} { \sigma(SM)\times Br_i(SM)}$. In
Figs.3-6, taking $m_h=125$ GeV, we show $\mu_{\gamma\gamma jj}$,
$\mu_{Vbb}$, $\mu_{\gamma\gamma}$, $\mu_{VV}$  $(V = W, Z)$ versus
the parameter $f$ respectively for the production channels
$qq'\rightarrow hjj \rightarrow \gamma \gamma jj$, $qq' \rightarrow
Vh \rightarrow Vbb$, $gg\rightarrow h \rightarrow \gamma\gamma$, $gg
\rightarrow h \rightarrow VV$ in Case A and Case B.
\begin{figure}[htbp]
\includegraphics[width=11cm]{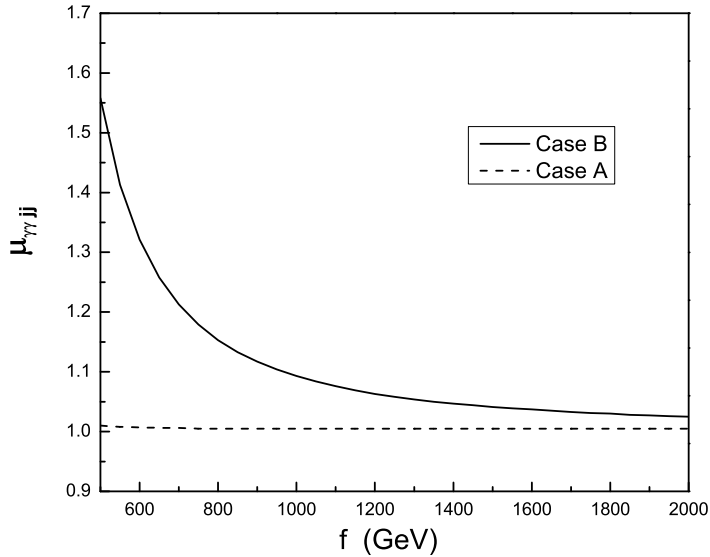}
\caption{\label{fig1}   The   $\mu_{\gamma\gamma jj}(qq'
\rightarrow jjh \rightarrow jj\gamma \gamma )$ rates as a function of
scale parameter $f$.}
\end{figure}
It is found that $\mu_{\gamma \gamma jj}$ corresponding to the
vector-boson fusion production followed by di-photon decay is larger
than one in the small $f$ region in Case B as shown in Fig.3. This was
also noted in Ref.~\cite{LHTrr}. This is mainly because in Case B,
both $h\rightarrow b\bar{b}$ and $h\rightarrow \tau\bar{\tau}$ are
significantly suppressed due to the shift in the couplings (as shown
in Eq.~7) which induces the branching ratios of $h\rightarrow
\gamma\gamma$ to increase. The increase effects dominate over the
reduced effects in production. The signal strength $\mu_{\gamma
\gamma jj}$ normalized to the SM prediction can reach $1.09-1.56$ for
parameter $f$ in the range $1000$ GeV$\sim 500$ GeV. The hint of
enhanced photon production rate in the vector boson fusion process
\cite{CMSVBF} can be interpreted as the effect of small $f$ in Case
B of the LHT model. The ratio for channel $qq' \rightarrow Vh
\rightarrow \gamma \gamma$ is nearly the same as $\mu_{\gamma \gamma
jj}$ because they mainly depend on the $VVh$ coupling and the
branching ratio of $h \rightarrow \gamma \gamma$. However, the
$\gamma\gamma$ enhancement can not be interpreted in Case A. If more
data are collected and it is further verified that events in the $\gamma
\gamma jj$ channel are indeed larger than the SM prediction, then  Case A
will be ruled out and   Case B will suffer further tests.
\begin{figure}[htbp]
\includegraphics [width=10cm] {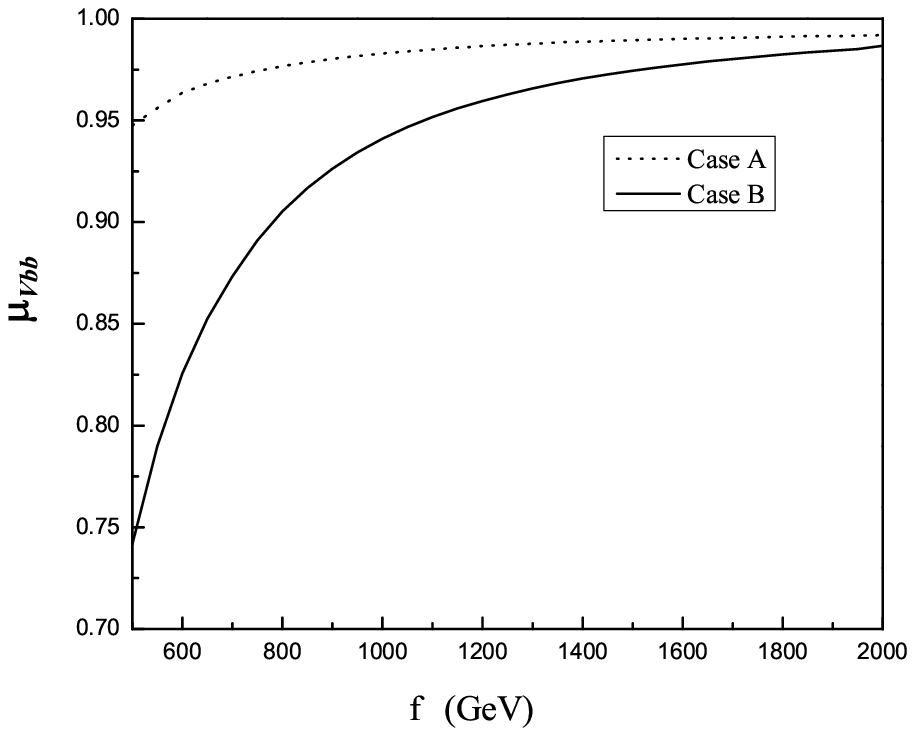}
\caption{\label{fig1}   The  $\mu_{Vbb}(qq' \rightarrow Vh
\rightarrow Vb\bar{b})$  rates as a function of scale parameter $f$.}
\end{figure}
\begin{figure}[htbp]
\includegraphics [width=10cm] {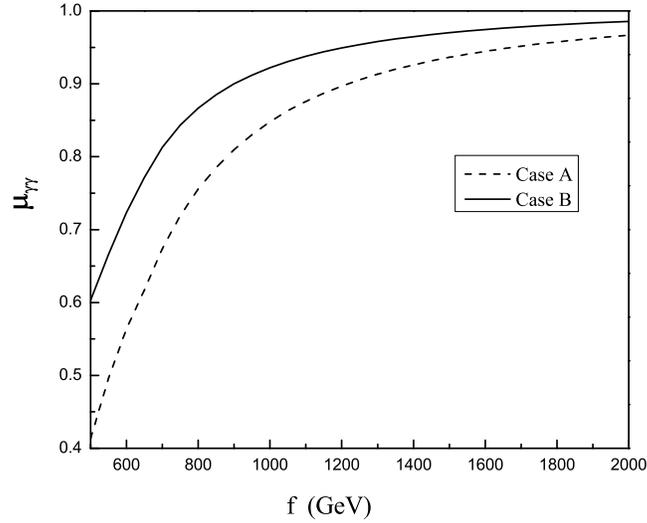}
\caption{\label{fig1}   The
$\mu_{\gamma\gamma}({gg\rightarrow h\rightarrow \gamma\gamma)}$ rates as a function of  scale parameter $f$.}
\end{figure}
The  $\mu_{\gamma\gamma}$ and $\mu_{VV}$  rates as a function of scale parameter  $f$
are also illustrated in Figs.~5-6. The ratios $\mu_{\gamma\gamma}$
and $\mu_{VV}$ are suppressed in both Case A and Case B because the
gluon-gluon fusion process is significantly suppressed by the
contributions from additional heavy fermions. The cross section for
subprocess $gg\rightarrow h$ can be represented as
\begin{equation}
\hat{\sigma}(gg\to h)=\Gamma(h\to gg)\frac{\pi^2}{8m^3_h}.
\label{gghhgg}
\end{equation}
 Both the T-even top partner and extra T-odd
fermions are considered in the loop calculation. The effective tree
level approximation in the so-called  heavy top limit is not used
\footnote{There are detailed discussions about the comparison
between loop calculation and heavy top approximation for the gluon-gluon
fusion process in Ref.~\cite{Colliderbook,ggLL}.}. The deviations
from the SM prediction for $\mu_{\gamma\gamma}$ and $\mu_{VV}$ are
 sensitive to the parameter $f$. When $f$ increases, the suppression is
  weakened sharply and the results are close to the SM predictions in
  the decoupling limit. Besides, the rate $\mu_{\gamma\gamma}$ for Case B is larger than that for Case A.
The main reason for this is that the large shift in $hb\bar{b}$ coupling
induces the significant increase of the branching ratio Br($h
\rightarrow \gamma \gamma$). A similar conclusion of the deviation
also holds for $\mu_{VV}$ in Fig.6.
 Our results are consistent with those in Ref.\cite{LHTrr,LHTrr2,LHTrr3}.
\begin{figure}[htbp]
\includegraphics[width=11cm]{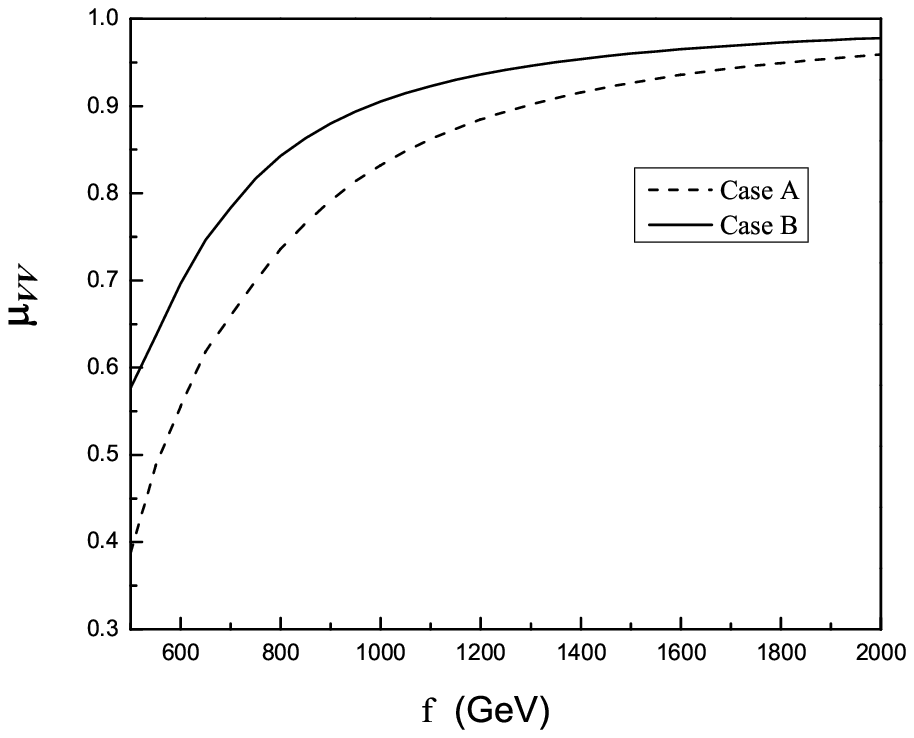}
\caption{\label{fig1}   The $\mu_{VV}(gg\rightarrow h
\rightarrow VV)$ rates  as a function of  scale parameter $f$.}
\end{figure}
As shown above, an enhancement factor of $1.09-1.56$ in the $qq'
\rightarrow jjh\rightarrow jj\gamma\gamma$ channel can be obtained
for the LHT model in Case B with parameter $f$ in the range $1000$
GeV$\sim 500$ GeV. And in the same region, the event rates in the $WW$
and $ZZ$ channels are slightly suppressed. Although the
$h\rightarrow b\bar{b}$ and $h\rightarrow \tau\bar{\tau}$ are
significantly suppressed, these two channels are not sensitive due
to the large backgrounds and relative low identification
efficiencies of the final states. In addition, the Higgs have not been
conclusively observed in these two channels. So, the Higgs boson of
the  LHT model in Case B with a small $f$ value could fit well
with the current Higgs search data. In Ref.\cite{Hfit2}, they got a
90\% CL favored region corresponding to low mass $m_T$ in the toy
LHT model.\footnote{Their study was carried out in the fermionic top
partner frame and only the most basic features of the LHT model are
kept under some assumptions. In particular, the $hVV$ couplings and
$hbb$ couplings are assumed to be consistent with those in SM.}
Based on the above calculation and analysis, we conclude that the
current Higgs search result favors the LHT model in Case B with a
low scale parameter $f$. The study in the framework of varying
Yukawa couplings \cite{XWan} also supports this conclusion. It is
expected that the Higgs sector of the  LHT model could be further
tested by the LHC  by increasing the integrated luminosity.

\section{Conclusions}
 \noindent

In both ATLAS and CMS analyses the decay channel $ h \rightarrow
\gamma\gamma$ has played an important role in discovering the Higgs
boson.
Motivated by the recent results from the Higgs search at the LHC,
extensive studies have been conducted to accommodate the hint of
$h\to \gamma\gamma$ enhancement
\cite{Hfit1,Hfit2,Hfit3,XWan,1207.3698,WTC,1201.5582,Scalar,THDM}.
 In general, the enhancement factor in $h\to \gamma\gamma$ can be
obtained via the contributions from new particles in the loop of
$h\to \gamma\gamma$ or by suppressing the Yukawa couplings in the
fermion sector. For the supersymmetric (SUSY) models, the
$h\gamma\gamma$ can be enhanced by the contribution of
$\tilde{\tau}$.  A comparative study on different SUSY models was
carried out in Ref.\cite{1207.3698}. This  shows that the most favored
SUSY model is the next-to-minimal supersymmetric model (NMSSM),
whose predictions about the 125 GeV Higgs boson can agree with the
experimental data at the $1\sigma$ level without any fine tuning
\cite{1207.3698} while the minimal supersymmetric standard model
(MSSM) suffers from some fine tuning \cite{1207.3698}.
  Ref.\cite{WTC} shows that the 125 GeV techni-dilaton in the walking technicolor (WTC)
  can be consistent with the current Higgs search results, where a large diphoton event rate can be achieved due to the
loop contributions of extra techni-fermions. In addition, the
minimal model of universal extra dimensions (MUED) explains how the
cross-sections for Higgs production via gluon fusion and decay into
photons are modified by KK particles running in loops
\cite{1201.5582}. On the other hand, in some other studies, the
enhancement factor in $h\to \gamma\gamma$ can be obtained by
modifying the Yukawa couplings \cite{XWan,THDM}.

In this paper, we have studied the Higgs production  and decay  of
the  LHT  model in the light of recent Higgs searchs at ATLAS and
CMS. The decay channels and cross sections for the 125 GeV Higgs
boson in  the LHT  model are presented. We found that the total
widths are suppressed in both Case A and Case B of the LHT model.
However, the branching ratios of $h\rightarrow \gamma\gamma$ are
enhanced in Case B in the small $f$ region since the main decay channel
$h\rightarrow b\bar{b}$ is significantly suppressed. The signal
rates normalized to SM prediction for Higgs ( $m_h=125$ GeV ) search
channels
 $qq' \rightarrow jjh \rightarrow jj\gamma \gamma$, $qq' \rightarrow Vh \rightarrow Vb\bar{b}$, $gg \rightarrow \gamma \gamma$,
$gg \rightarrow WW$, $gg \rightarrow ZZ$ are also presented. It is
found that an enhancement factor of $1.09-1.56$ in the $qq' \rightarrow
jjh\rightarrow jj\gamma\gamma$ channel can be obtained for the LHT
model in Case B with parameter $f$ in the range $1000$ GeV $\sim
500$ GeV. In the LHT model, the rates for $b\bar{b}$,
$\tau\bar{\tau}$ in Case B are significantly suppressed relative to
the  SM predictions that are still consistent with the current statistic. It is expected that the Higgs properties of
  the  LHT model could be further tested with a larger data sample at the LHC.

\vspace{4mm} \textbf{Acknowledgments}\\
We would like to thank Qi-Shu Yan and Xia Wan for helpful discussions.This work was supported in part by the National Natural Science Foundation of
China under Grants No.10975067 and 11175251, the Postdoctoral Science Foundation of China (No.2012M510248),
the Natural Science Foundation of the Liaoning Scientific Committee
(No. 201102114), and Foundation of Liaoning Educational Committee (No. LT2011015).

\end{document}